\documentclass{aa}

\usepackage{graphics}
\usepackage{psfig}

\begin{document}

\title
{Tidal dwarfs in the M81 group~: the second generation?
\thanks{Based on observations with the NASA/ESA
Hubble Space Telescope, obtained at the Space Telescope Science
Institute, which is operated by the Association of Universities for
Research in Astronomy, Inc., under NASA contract NAS 5-26555. These
observations are associated with proposal ID GO-8192.}}

\author{L. N. Makarova\inst{1,2}
	\and
	E. K. Grebel\inst{3}
	\and
	I. D. Karachentsev\inst{1}
	\and
	A. E. Dolphin\inst{4}
	\and
	V. E. Karachentseva\inst{5}
	\and
	M. E. Sharina\inst{1,2}
	\and
	D. Geisler\inst{6}
	\and
	P. Guhathakurta\inst{7,8}
	\and
	P.W. Hodge\inst{9}
	\and
	A. Sarajedini\inst{10}
	\and
	P. Seitzer\inst{11}}

\institute{Special Astrophysical Observatory of the Russian Academy of Sciences,
 Nizhnij Arkhyz 369167,
Karachaevo-Cherkessia, Russia\\
\email{lidia@sao.ru, ikar@luna.sao.ru}
\and
Isaac Newton Institute of Chile, SAO Branch
\and
Max-Planck Institute for Astronomy, K\"onigstuhl 17, D-69117 Heidelberg, Germany
\and
Kitt Peak National Observatory, National Optical Astronomy Observatories,
PO Box 26732, Tucson, AZ, 85726, USA
\and
Astronomical Observatory of Kiev University, 04053, Observatorna 3, Kiev,
Ukraine
\and
Departamento de Fisica, Grupo de Astronomia, Universidad de Concepci\'on, Casilla 160-C,
Concepci\'on, Chile
\and
Herzberg Fellow, Herzberg Institute of Astrophysics, 5071 W.\ Saanich Road,
Victoria, B.C.\ V9E~2E7, Canada
\and
Permanent address: UCO/Lick Observatory, University of California at Santa
Cruz, Santa Cruz, CA 95064, USA
\and
Department of Astronomy, University of Washington, PO Box 351580, Seattle, WA 98195, USA
\and
Department of Astronomy, University of Florida, Gainesville, FL 32611, USA
\and
Department of Astronomy, University of Michigan, 830 Dennison Building, Ann Arbor, MI 48109, USA
}

\titlerunning{Tidal dwarfs in the M81 group}

\authorrunning{Makarova et~al.}

\abstract{ We derive quantitative star formation histories
of the four
suspected tidal dwarf galaxies in the M81 group, Holmberg~IX, BK3N,
Arp-loop (A0952+69), and Garland, using Hubble Space Telescope/Wide Field
Planetary Camera~2 images in F606W and F814W obtained as part of a Snapshot
survey of dwarf galaxies in the Local Universe.
We consider the spatial distribution and ages of resolved stellar populations
in these dwarf irregular galaxies.  We use synthetic color-magnitude
diagrams to derive the ages of the major star formation episodes, star
formation rates, and approximate metallicity ranges.  
All the galaxies show evidence of continuous star formation between about
20 and 200~Myr ago with star formation rates in the range
$7.5\cdot10^{-3}$--$7.6\cdot10^{-4}$~M$_{\sun}$~yr$^{-1}$.
The metallicity of the detected
stars spans a wide range, and have lower than solar abundance.
A possible scenario is that all four dwarf galaxies were formed from 
material in the metal-poor outer part of the giant spiral galaxy M~81
after the tidal interaction between M~81, M~82, and NGC\,3077 $\sim 200$~Myr
ago.  While we do not directly detect pronounced old
stellar populations, the photometric limits of our data are such that the
presence of such a population is not entirely ruled out.
}

\maketitle

\section{Introduction}

Galaxy interactions are still common at the present epoch and 
can have a significant impact on the appearance and
evolution of galaxy groups and clusters.  The closest examples of 
ongoing interactions can be found in our immediate surroundings:
The Milky Way is currently accreting the Sagittarius dwarf
spheroidal (dSph) galaxy (Ibata et~al.\ \cite{ibata}) and interacts with
the Magellanic Clouds (e.g., Gardiner, Sawa, \& Fujimoto \cite{gsf}).
Evidence for more dramatic interactions involving an entire galaxy
group can be seen in the nearby M81 group, whose dominant
members M81, M82, NGC\,3077, and NGC\,2976 are embedded in a
huge H\,{\sc i} cloud with extended tidal H\,{\sc i} bridges 
between the components (van der Hulst \cite{hulst};
Appleton et~al.\ \cite{appleton};
Yun, Ho, \& Lo \cite{yun}). Past surveys for dwarf galaxies in the M81
group led to the identification of a large number of dwarf galaxy
candidates (B\"orngen \& Karachentseva \cite{bk}; Karachentseva, Karachentsev,
\& B\"orngen \cite{kkb}; van Driel et~al.\ \cite{vandriel}; Boyce et~al.\ 
\cite{boyce}; Karachentsev, Karachentseva \& Huchtmeier \cite{k2001}).
Some of the concentrations visible in the H\,{\sc i} bridges 
have candidate optical counterparts identified as
dwarf galaxies.  It has been suggested that some of these dwarf
galaxies are tidally disrupted dwarfs (e.g., Karachentseva, Karachentsev, 
\& B\"orngen \cite{kkb}), while others may be dwarfs newly formed in the tidal
tails (e.g., van Driel et~al.\ \cite{vandriel}).

Owing to its proximity (mean distance 3.7 Mpc; Karachentsev et~al.\
\cite{karachentsev})
the M81 group offers a unique opportunity to study the properties of
tidal dwarf galaxy candidates in great detail.  Of particular interest
are the evolutionary histories of the potential tidal dwarfs and the
effect of tidal interaction on star formation.

In the framework of our Hubble Space Telescope (HST)
snapshot survey of nearby dwarf galaxy candidates (programs SNAP 8192
and 8601, PI: Seitzer)  we have imaged 25
galaxies in the M81 group (Karachentsev et~al.\ \cite{k99}, \cite{k00},
\cite{k01}, \cite{karachentsev}). Photometric distances
of these galaxies were obtained using the tip of red giant branch (TRGB),
a well-established distance indicator (e.g., Lee et~al.\ \cite{lee}).

Here we present an analysis of data on four dwarf irregular galaxies located
in the tidal H\,{\sc i} features of the M81 group (Holmberg~IX,
BK3N, Arp-loop and Garland). In Section 2 we
describe the observations, reduction and photometric errors analysis.
In Section 3 we discuss the resulting color-magnitude diagrams.
In Section 4 a quantitative analysis of star formation in the galaxies
is given. Section 5 contains a general discussion of star formation history
and evolution of the galaxies, as well as the conclusions.

\section{Observations and data reduction}

The galaxies were observed with the Hubble Space Telescope during 2000 July
30 --- 2001 June 27 with the Wide Field and Planetary Camera (WFPC2)
as a part of the Dwarf Galaxy Snapshot Survey (Seitzer et~al.\ \cite{seitzer},
Grebel et al.\ \cite{grebel}).
For each target, two exposures of 600 sec each were obtained, one in the
F606W filter and one in the F814W filter.

Stellar photometry of the galaxies was carried out by A.\ E.\ Dolphin 
with the HSTphot program (Dolphin \cite{dolphina}).
Before running photometry, the data quality images were used to mask bad
pixels ({\it mask} procedure of the HSTphot package). Bright cosmic rays were
then masked out using the HSTphot {\it cleansep} utility.
The stellar
photometry was obtained simultaneously on the F606W and F814W frames with
the {\it multiphot} procedure. HSTphot uses a library of model PSFs based
on Tiny Tim. Changes of the focus of HST were corrected using the residuals
of bright, isolated stars. Resulting instrumental magnitudes of 0$\farcs$5
radius were corrected for charge-transfer inefficiency and calibrated
using the relations of Dolphin (\cite{dolphinb}). We estimate the uncertainty
of the photometric zeropoint to be within 0$\fm$05 (Dolphin \cite{dolphinb}).
Only stars of better photometric quality were used for further analysis.
We selected stars with photometric errors less than 0.2 mag,
$\chi \leq 2$, and $-0.3 \leq$~sharpness~$\leq$0.3 in both filters.
Photometry results for the galaxies are presented in Table~1, which is
available only in electronic form in the CDS. This table contains
1650 stars measured in the Holmberg~IX field, about 600 stars in the BK3N field,
about 4700 stars in the Garland field and about 250 stars in the Arp-loop field.

Data processing and photometric reduction procedures were identical to
the ones used in our previous studies of nearby dwarf galaxy candidates
(see, for example, Dolphin et~al.\ \cite{d2001}).

Artificial star tests were performed for each of the galaxies using the same
reduction procedures. These tests allow us to estimate the accuracy
and completeness of our photometry (see Fig.~1). As was noted earlier
(Dolphin et~al.\ \cite{d2001}), the $\sim$ 90\% maximum completeness
level is primarily a result of the fraction of stars falling within
one pixel of a masked pixel (cosmic ray, bad pixel, bad column, etc.)
and are thus rejected.

\section{Color-magnitude diagrams}

\subsection{Distance and reddening}

Color--magnitude (V--I,I) diagrams of the four galaxies under consideration
are presented in Fig.~2.
Galactic extinction is not large for these galaxies: $A_I=0\fm15$ for
Holmberg~IX, $A_I=0\fm16$ for BK3N and Arp-loop and $A_I=0\fm13$ for
Garland, according to Schlegel et~al.\ (\cite{schlegel}). The Galactic
extinction correction was not made
in Fig.~2 in order to facilitate the comparison with the
M~81 CMD below (Fig.~3), which also was not
corrected for Galactic extinction. Isochrones of appropriate
metallicities and ages from
Girardi et~al.\ (\cite{girardi}) are shown in the color-magnitude
diagrams (CMDs). The isochrones were shifted by the appropriate Galactic
extinction and the galaxies' distance moduli.
The M~81 Cepheid distance modulus ($\mu_0$ = 27.80~mag,
Freedman et~al.\ \cite{freedman}) was taken for the three galaxies
Holmberg~IX, BK3N, and Arp-loop.
For Garland we used the 
distance modulus obtained from the tip of the red giant branch
(TRGB) of the NGC~3077 halo population (Karachentsev et~al.\ \cite{karachentsev}).
BK3N and Arp-loop also have TRGB distance estimations (which agree within
errors with the Cepheid distance, see
Karachentsev et~al.\ (\cite{karachentsev})), but whether the detected red
giants belong to the galaxies is unclear. We also cannot
suggest a significantly larger distance for the objects since they are 
clearly detected in the core M~81 H\,{\sc i} complex.
The metallicity specified in each diagram is the metallicity of
the theoretical
isochrones shown according
to our best-fit models. We did not estimate a metallicity from
the color and slope of the red
giant branch stars, because these stars are too close to the 
photometric detection limit.

The color-magnitude diagrams in Fig.~2 are quite typical for dwarf
irregular galaxies. We can see the upper main sequence and probable
helium-burning blue loop stars,
the red supergiant branch and probably also some young and intermediate
age AGB stars for all of the galaxies. There are no clear signs of an RGB
in the Holmberg~IX CMD, as was already noted
by Karachentsev et~al.\ (\cite{karachentsev}). The apparent absence
of an old stellar population may be direct evidence of the recent
formation of Holmberg~IX (see next section for details). In contrast 
to Holmberg~IX, Garland shows a clump of stars at faint magnitudes
that consists very likely of RGB stars.  These, however, may belong to the
nearby NGC~3077 galaxy, not to Garland itself (see also
Sakai and Madore (\cite{sm})). The other two irregular dwarfs
show a slightly increased number of stars at $I>24$ mag, which may be
indicative of the tip of their respective red giant branches.

\subsection{Foreground/background contamination}

As we can see from Fig.~2,
the color-magnitude diagrams of the suspected tidal dwarfs are not heavily
contaminated by foreground stars.
The expected number of Galactic foreground stars
in the magnitude range 18\fm0~$\le$~I~$\le$~20\fm0 would be
roughly two or three in the WFPC2 field according to the star counts
of Bahcall and Soneira (\cite{bs}).
We have made also independent number counts of the stars in the CMDs of 
24 observed dwarf galaxies of the M~81 group in the same magnitude range
18\fm0~$\le$~I$_0$~$\le$~20\fm0
and in the color range 1.4~$\le$~(V-I)$_0$~$\le$~3.0. 
We included data for M~81 group dwarf irregulars
as well as dwarf spheroidals in these counts.  
(NGC~2366 was excluded, because it contains 22
bright stars in this area.) The mean number 
of stars fulfilling these selection criteria is 2.4 with a standard 
deviation of 2.4.  Confining ourselves to the nine dwarf spheroidals
in the M~81 group for which we have HST data, we find stellar number
counts of 1.2$\pm$1.5 in the afore described range.
These numbers are in good agreement with the Bahcall \& Soneira (\cite{bs})
model. Number counts for fainter magnitudes may be significantly contaminated
by red supergiant branch stars of the irregular galaxies.  The number of stars
in the magnitude range 18\fm0~$\le$~I~$\le$~22\fm0 and in the color range
1.4~$\le$~(V-I)$_0$~$\le$~3.0 for dwarf spheroidal
galaxies is 3.4$\pm$2.
Therefore, we expect between 1 to 5 foreground stars in the
magnitude range 18\fm0~$\le$~I~$\le$~22\fm0 and in the color
range 1.4~$\le$~(V-I)$_0$~$\le$~3.0 in WFPC2 field.

The color-magnitude diagrams of Holmberg~IX, BK3N, and Arp-loop may also
be contaminated by stars belonging to the close spiral galaxy M~81.
For comparison we show a V--I,I CMD of M~81 in Fig.~3.
This diagram was taken from Hughes at al. (\cite{hughes}).
It shows a field along M~81's major axis at an angular distance from
the nucleus of about 5$\farcm$5. The observations were made with
the Wide Field Camera (WFC)
aboard the Hubble Space Telescope as a part of the Extragalactic Distance
Scale Key Project. The color-magnitude diagram in Fig.~3 was not
corrected for Galactic extinction. We can see upper main sequence
stars, red supergiant branch (RSGB) stars, and red giant branch stars
in this diagram.
The blue plume and RSGB  are very wide.  The RGB falls in the region of large
photometric errors. As was noted by
Hughes at al. (\cite{hughes}),
the color spread observed in M~81 is probably due to a combination of
photometric errors and variable reddening across the field. The red supergiant
population seems also to be very red in general. Based on H\,{\sc ii} region
observations, the mean abundance of recently formed stars 
is 12+log(O/H)=8.85 (i.e., about the solar abundance) for this field
(Freedman et~al.\ \cite{freedman}).
Note that Galactic extinction is nearly the same for
M~81 and for the three dwarf irregulars.
The color-magnitude diagram of M~81 is $\sim$1$\fm$5 deeper in I than our
CMDs and it contains ten times more stars.
As can be seen from the comparison of the CMDs in Fig.~2 and Fig.~3,
the relatively well-populated RSGB of Holmberg~IX
has slightly bluer colors than the majority of M~81's red supergiants.
The red supergiant branch of Holmberg~IX is also considerably more narrow
than in M~81, which appears to indicate low internal extinction (in addition
to a possibly lower metallicity). The poorly populated RSGBs
of BK3N and Arp-loop generally agree with the location of  
Holmberg~IX's RSGB in the CMDs. Also, the blue features in the
diagrams of the dwarfs and in M~81's CMD have the same overall location.
The red giant branch of M~81 in Fig.~3 appears to extend farther to redder
colors by at least $\sim 0.5$ mag 
than observed in our dwarfs, which may indicate that a number of
red giants in M~81
have a higher metallicity than the few red giants in the dwarf galaxies. 
On the other hand, we cannot exclude reddening effects introduced by 
variable extinction intrinsic to M~81, and effects of photometric errors
on this close to the photometric limit area of the CMD.
Also, we can not exclude contamination
of our CMDs by M~81 red giants, if M~81 has an
extended old halo population.

All of our data may be contaminated by background galaxies and quasars.
While the excellent resolution of HST's WFPC2 allows us to reject many
galaxies based on their extended light profiles, a few quasars may
contribute to the point-source number counts at blue colors.

\section{Star formation history of the tidal dwarfs}

It is well known that the central galaxies of the M~81 group are strongly
interacting. There is a huge cloud of intergalactic H\,{\sc i} encompassing
M~81, M~82, NGC~3077 and NGC~2976 (Appleton et~al.\ \cite{appleton},
Yun et~al.\ \cite{yun}). The image of the hydrogen complex from the paper
of Yun et~al.\ (\cite{yun}) is shown in Fig.~4. This strong interaction
may be a reason for the formation of tidal dwarf galaxies from a condensed
interstellar medium that was stripped from the main body of one of the galaxies
in the interacting system.

Recently results of a blind H\,{\sc i} survey of the M~81 group were published
by Boyce et~al.\ (\cite{boyce}). It was suggested in this article that
Holmberg~IX and Arp-loop may have recently condensed from the tidal
debris between M~81 and M~82 (see Fig.~4).
The origin of Garland is uncertain, although it is plausibly
associated with the tidal interaction of NGC~3077 with M~81 (Karachentsev,
Karachentseva \&Borngen \cite{kkb}).
Recently Heithausen \& Walter (\cite{hw}) revealed a giant molecular
cloud in the Garland region with a mass of about 3~$\times$~10$^7$~M$_{\sun}$.
Van Driel et~al.\ (\cite{vandriel})
suggested that Garland may be at an intermediate stage in the conversion
of a tidal tail into a dwarf galaxy. It was noted by
Boyce et~al.\ (\cite{boyce}), that the origin of BK3N is rather controversial.
The object may be
a tidal dwarf galaxy, condensing from the tidal debris of the spur.
Alternatively, BK3N may be a pre-existing dIrr galaxy undergoing an
interaction with M~81.

Below we attempt to make a quantitative measurement of the star formation
history of Holmberg~IX, BK3N, Garland and Arp-loop. We consider the spatial
distribution and ages
of the resolved stellar populations in these possible tidal
dwarfs in some detail and try to elucidate the origin of these galaxies.

\subsection{The method}

We used the {\bf starFISH} package by Harris \& Zaritsky (\cite{hz})
for the star formation history analysis. This package is intended to determine
the best-fit star formation history (SFH) of a mixed stellar population.
The package constructs a library of synthetic CMDs based on theoretical
isochrones, and determinations of the interstellar
extinction, photometric errors, and distance modulus based on the
observational data. These synthetic
CMDs are combined linearly and compared to the observed CMDs using
$\chi^2$ statistics. 
We used theoretical isochrones from the Padua group (Girardi et~al.\ 
\cite{girardi}).  To include also isochrones with Z=0.001 and log(Age)$<$7.8
we complemented this set by earlier work of the same group
for these ages and metallicity (Bertelli et~al.\ \cite{bertelli}).
Merging these sets did not affect our 
calculations, because the isochrones of the  neighboring metallicities
Z=0.0004 and Z=0.004 show considerable differences only
for stars with masses greater than 40 M$_{\sun}$. Such massive stars are not
present in our data.

Each isochrone gives us absolute magnitudes and color indices
(including Johnson-Cousins V and I) of a stellar population with
a particular age and metallicity. The program transforms these magnitudes
into a synthetic CMD by accounting for the following input parameters~:
galaxy distance, interstellar
extinction, initial mass function (IMF), binary fraction, and photometric
errors. We constructed the CMDs from isochrones of the full available age range
(from $10^{6.6}$~yr to $10^{10.2}$~yr) with a step of log(Age)=0.2.
We are not aware of any metallicity estimations from other studies of 
Holmberg~IX, BK3N, and Arp-loop. The metallicity of red giant stars
in the halo of NGC~3077 near the Garland region was estimated photometrically
by Sakai \& Madore (\cite{sm}) from their HST/WFPC2 CMD to be Z=0.005.
Therefore, for all the galaxies a number of fits were made with
different metallicity combinations using the entire range of the available
metallicities from Z=0.0004 to Z=0.019. The best solutions were then
chosen taking into account $\chi^2$--goodness-of-fit parameter and
also the general agreement between the observed CMD and the synthetic CMD
constructed from the star formation history (SFH) solution.
We have no information about internal extinction in the dwarf galaxies.
Therefore, some experiments with differential extinction values for young
and old stars were made. The introduction of additional internal extinction
for young stars, which mainly populate the CMDs, gave us an RSGB that was too wide
and scattered and also some additional scatter in the blue features.
As a result, only Galactic extinction was used for the calculation.
Detailed photometric errors and completeness data from our artificial
star tests were also inserted. We used the default values for the IMF
(the Salpeter law) and the binary fraction equal to 0.5.
Each resulting model CMD represents a stellar population with the age and
metallicity of the parent isochrone. A linear combination of these CMDs
forms a composite model CMD that can represent any SFH. The amplitude
associated with each CMD is proportional to the number of stars formed
at that age and metallicity. The best-fit SFH is described by the set
of amplitudes that produces the composite model CMD most similar to
the observed CMD. The best-fitting amplitudes are determined by using
a modified downhill simplex algorithm (see Harris \& Zaritsky \cite{hz}).

Our photometry of the stars in the dwarf galaxies contains only
stars sufficiently luminous to be detectable in 10-min exposures with HST/WFPC2.
The number of stars in the CMDs is low (about 250 in Arp-loop, for example).
These factors make the calculations especially difficult, and uncertainties
in the SFR estimation are rather large (see below). Nevertheless,
most of the galaxies have goodness-of-fit $\chi^2 <$~10. These are quite good
results according to J.~Harris (starFISH User's Guide).

\subsection{Holmberg~IX}

There are no clear signs of the red giant branch in the CMD of
this object, as was already noted by Karachentsev et~al.\
(\cite{karachentsev}) (see Fig.~2).
The resolved stellar populations are mainly blue and red supergiant stars.
These two branches are relatively well populated, which distinguishes 
Holmberg~IX from the other three galaxies.
The spatial distribution of the various populations in Holmberg~IX
is shown in Fig.~5. The stars in the area (V--I)$_0$~$\ge$0.8~mag and
23\fm75~$\le$~I$_0$~$\le$24\fm5 are distributed rather homogeneously across 
the field. They do not show any concentration towards the galaxy body
and may be old red giants belonging to M~81. Upper main sequence
stars (MS)
are concentrated clearly in the visible star formation areas of
the galaxy. The blue loop and red supergiants show a clear increase in 
density toward the central regions of Holmberg~IX.
Only the upper, luminous part of the blue population 
($I_0$~$\le$23\fm75) is shown
in Fig.~5.  When all the blue stars in Holmberg~IX's CMD ($I_0$~$\le$24\fm5)
are plotted instead, the central concentration becomes less pronounced,
indicating that recent star formation was indeed strongly
centrally concentrated.  For a general discussion of population gradients
in dwarf galaxies, see Harbeck et~al.\ (\cite{Harbeck}).
A full mapping of the more extended, fainter/older populations requires
a larger field of view than the partial coverage of {Holmberg~IX
afforded by our HST/WFPC2 pointing.

The best-fit SFH for Holmberg~IX is shown in Fig.~6. For this model
the $\chi^2$ value is 8.1. The model CMD constructed
from this SFH is shown at the right panel of Fig.~7.
 For comparison the original CMD
of the galaxy is displayed in the left panel of Fig.~7. As can be seen
from the comparison of these two panels, there are differences
between the diagrams. The observed groups of red and
blue stars are generally slightly redder (about 0.1 - 0.2 mag in V--I).
The differences between observed and model data are most clear in the RSGB
slope, which is somewhat steeper for the model branch. Introducing
additional extinction does not improve the solution.  The red
supergiant branch is sensitive especially to this internal extinction,
which produces too much scatter as compared to the observed RSGB.
The model CMD in Fig.\ 7 contains fewer bright blue and red supergiants.
This lack is common for the model CMDs of all four dwarfs.  The
very small number of the brightest stars in the original data makes statistical
analysis and modelling especially difficult in this part of the diagram.
This may be the main reason for the deficiency of the brightest
stars in the model diagram relative to this population of the original data.
A number of red giants, which probably belong to M~81, are
also reproduced in Fig.~7. Their age was estimated to be about 2--8 Gyr.
However, the presence of these stars
cannot give noticeable enhancement of ancient star formation in our
model (see Fig.~6).
Another, more general problem affecting any population modelling is the 
ability (or inability) of theoretical isochrones to fully account for all
features in an observed CMD (see, e.g., Langer \& Maeder \cite{LanMaed}
for red and blue supergiants).

We find a dominant episode of star formation with an age range from about
6 to about 200~Myr. The mean star formation rate (SFR) during this period
was 7.5$^{+2.6}_{-2.2}\cdot$10$^{-3}$~M$_{\sun}$~yr$^{-1}$ with a metallicity 
spanning $-0.7 \leq$[Fe/H]$\leq-0.4$. The wide range in metallicities can
indicate that the method is unable to resolve
metallicity differences of $\leq$0.3 dex in our data.
For instance, differences in the location of main sequence stars caused
by metallicity within this range become indistinguishable in the presence
of differential reddening.
There are no clear signs of older star formation although, given the limitations
of the data, we cannot exclude the presence of sparse older populations.

Thus, we can see continuous star formation in Holmberg~IX from about
6 until 200~Myr ago and probably also a small number of younger stars.
Therefore, Holmberg~IX may indeed be a very young
tidal dwarf galaxy formed out of tidal debris from the interaction between M~81
and M~82. An indirect
confirmation of the age estimation comes from results of a numerical
simulation of the dynamical evolutionary history of the M~81--M~82--NGC~3077
system (Yun \cite{y99}): The nearest approach between M~82 and M~81
was 220~Myr ago according to this work, which is in good agreement with
our result.

Spectroscopic observations of Holmberg~IX were carried out at the 6-m BTA
telescope (Special Astrophysical Observatory, Russia) by S.~A.~Pustilnik
and A.~G.~Pramsky using the fast spectrograph of the prime focus (see the equipment description
on {\tt http://www.sao.ru/Doc-en//Telescopes/bta/instrum/ instrum.html}). The slit was placed
on one of H\,{\sc ii} regions, which was found within the galaxy field by Miller \& Hodge
(\cite{mh}) (region number 8). This region is also visible in our WFPC2 field.
The spectrum of this object is rather weak, and only H$\beta$, H$\alpha$
and weak [NII]$\lambda$6584 are seen. Using an empirical relation between O/H
and the flux ratio of [NII]$\lambda$6584 and H$\alpha$ (van Zee et~al.\ \cite{vanzee};
Denicol\'o et~al.\ \cite{denicolo}), the derived 12+log(O/H) value is equal to 8.50 with probable
r.m.s. uncertainty of 0.15 dex (or [O/H]=--0.42). The latter corresponds to
an ionized gas metallicity of Z$\approx$0.0076.
This
value falls within the stellar population metallicity range found in our modelling.

\subsection{BK3N}

This small galaxy is located at the southwestern side of M~81, almost
symmetrically to Holmberg~IX at the opposite side of M~81 (see Fig.~4).
However, in distinction
to Holmberg~IX, signs of RGB stars in the CMD of BK3N can be recognized.
Other than that, mainly blue loop/blue main sequence stars and red 
supergiants are present in the diagram.

The spatial distribution of resolved stars in the object is shown in Fig.~8.
We can see the strong concentration of the upper main sequence stars
((V--I)$_0$~$\le$0) towards the galaxy. The blue loop also
exhibits a very clear concentration to BK3N, albeit with a larger spatial
extent.  The red supergiants
show a less concentrated distribution
and trace the body of the dwarf only marginally. In the right bottom
part of Fig.~8 the distribution of red giants is shown. We can distinguish
a marginal increase in the stellar density toward M~81 and absence of
any concentration of this population towards BK3N.  This seems to indicate
that the red giants in the CM diagram of this dwarf galaxy
belong to M~81.  Like Holmberg~IX, this makes BK3N another candidate
of a young tidal dwarf galaxy.

The small number of red and blue supergiants in the CMD of BK3N causes
very large uncertainties in the SFR determination, and the $\chi^2$ uncertainty for
the fitting is 2.3.
The result of the SFH computation for this galaxy is presented in Fig.~9.
The most significant star formation occurred in this galaxy
about 8--200~Myr ago, as can be seen from the histogram in Fig.~9. The star
formation rate in this period is estimated to be
3.8$^{+8.9}_{-1.7}\cdot$10$^{-4}$~M$_{\sun}$~yr$^{-1}$ with 
a metallicity range to be --0.7$\leq$[Fe/H]$\leq$0. The formally
allowable metallicity spread is thus even larger than for Holmberg~IX.
The large uncertainty in
the SFH determination does not allow us to draw reliable conclusions
about ancient star formation in this galaxy. Nevertheless, some star
formation may have taken place about 500~Myr--2~Gyr ago and 8--12~Gyr ago with
a probably lower metal abundance
([Fe/H] about --0.7). The original and model CM diagrams are presented
in Fig.~10. Like for the previous galaxy, we can see some deficiency of
bright red and blue stars in the model CMD, but the main features in the diagram
correspond to the observed populations, and
red giants, which probably belong to M~81, are also reproduced
in our model (see Fig.~10).
However, the data quality for this object is so poor that modelling 
will remain ambiguous.  Deeper data are clearly needed.

The potential presence of the intermediate-age stars (500~Myr--2~Gyr)
in the galaxy may indicate that BK3N formed 
before the tidal interaction in the core M~81 group.
However, these stars may also have been stripped from M81's outer parts during
the interaction.

\subsection{Arp-loop (A0952+69)}

This object is the brightest part of the diffuse circular structure
that embraces the northern part of M~81.
Our observations of this galaxy at the 3.5-m telescope of Apache Point
Observatory (USA) show that Arp-loop has an extremely low surface brightness,
which is fainter than
26 mag~arcsec$^{-2}$ in the V band (Makarova et~al.\ \cite{makarova}).

The color-magnitude diagram of Arp-loop contains the smallest number of stars
(about 250) among the four galaxies studied (see Fig.~2).

The spatial density distribution for the resolved stars in the galaxy CMD
is shown in Fig.~11. There are only a few objects with (V--I)$_0$~$\le$0,
which may belong to the upper MS of Arp-loop, or which could in part 
be background point sources. The distribution of blue loop stars 
shows a clear concentration toward the center of the galaxy. 
These stars trace star formation regions in the
galaxy very clearly, as do the RSG stars, which are similarly concentrated.
Red giants also smoothly trace
the galaxy overall structure and exhibit a clear concentration toward the
center of Arp-loop, which supports the idea that they belong to
Arp-loop rather than to M~81.
However, the location of these stars in the WFPC2 field coincides with
the direction toward M~81.
It is conceivable that one can see the northern part
of M~81's outer edge of red giant disk.

The best-fit SFH for this galaxy is shown in Fig.~12.
For this model $\chi^2$ value is 1.5.
We find a dominant
episode within an age interval of about 40--160~Myr ago and a 
metallicity spanning a range of [Fe/H] = $-1.3$ dex to $-0.4$ dex.
We obtain the mean
star formation rate of 3.6$^{+9.7}_{-2.1}\cdot$10$^{-4}$~M$_{\sun}$~yr$^{-1}$.
A less intensive star formation episode may have taken 
place about 500~Myr to 1~Gyr ago.
One can also see signs
of earlier star formation about 5 to 8~Gyr ago with a SFR of
2.2$^{+5.2}_{-1.6}\cdot$10$^{-4}$~M$_{\sun}$~yr$^{-1}$
and a metallicity spanning a range of [Fe/H] = $-1.3$ to $-0.7$ dex.
The photometric limit does not allows us to detect
older stars in this galaxy.
The resulting model CM diagram for Arp-loop is shown in the right panel of
Fig.~13. As we can see from the comparison with the observational data (left
panel in Fig.~13), there is a good agreement of these two diagrams.

Thus one possible conclusion from our data is that Arp-loop may be
a pre-existing dIrr galaxy undergoing strong interaction with M~81.
On the other hand, red giants in Arp-loop show a clear concentration toward
the direction of M~81.  They are presumably at least in part constituents
of the outer part of M~81's red giant disk.  If all of the red giants
in fact belong to M~81, then this would make Arp-loop into a tidal dwarf
with an age of approximately 1~Gyr.  Alternatively, Arp-loop could be part
of an outer spiral arm of M~81 with very low surface brightness.

\subsection{Garland}

Garland is situated very close to NGC~3077, and all of the WFPC2 field
is filled with AGB/RGB stars that appear to belong to NGC~3077. The stellar
population of Garland was recently studied in some detail by Sakai and Madore
(\cite{sm}). Their observations were made with HST/WFPC2
using the F555W and F814W filters.  The
resulting CMDs are 1$\fm$5 deeper than ours. The authors obtained a TRGB
distance modulus of the galaxy of $\mu_0$ = 27.93 mag. The distance modulus
$\mu_0$ = 27.89 mag
found by Karachentsev et~al.\ (\cite{karachentsev}) is in good agreement
with the result of Sakai and Madore (\cite{sm}).

However, neither the data of Sakai and Madore, nor our data allow us
to confirm or rule out the presence of RGB stars in Garland itself.
Hence it is unclear whether the derived distance refers to NGC~3077,
to Garland, or to both.
We use theoretical isochrones of Z=0.004 to fit the color-magnitude
diagram of Garland, taking into account Sakai and Madore's (\cite{sm}) 
estimation
of the RGB metallicity (Z=0.005) from the colors of the RGB stars.
The isochrones fit the CMD quite well, except for some red stars with
(V--I)$_0$~$>$~2 and I$_0$~$<$~21, which may be Galactic foreground stars.

The spatial distribution of the resolved stellar population is shown in Fig.~14.
It clearly demonstrates the large number of NGC~3077 RGB stars. There is
also a population of upper main sequence stars with
(V--I)$_0$~$\le$~0 and I$_0$~$>$~22 which trace an arc-like structure
distinct from the distribution of the RGB stars.

Unfortunately, the algorithm used for the SFH calculation did not
result in an acceptable fit.
Therefore, we decided to omit a quantitative SFH analysis here.
In general, we can recognize in the CMD evidence for star formation in age
intervals similar to the two previous galaxies. Namely, recent episode
of star formation probably occurred about 16--200~Myr ago.

As noted by Sakai \& Madore, for the case of Garland,
it is conceivable that recent star formation (16--200~Myr ago) was
triggered as a direct result of the tidal interaction of the galaxy
in the core M~81 group. However, the overall age and nature of this feature
near NGC~3077 remain uncertain due to the presence of a large number of
NGC~3077 red giant stars in the Garland field.

\section{Discussion and conclusion}

We investigated the question of whether the four dIrr galaxies Holmberg~IX,
BK3N, Arp-loop, and Garland, whose projected positions coincide with 
density peaks in the H\,{\sc i} tidal streams in the interacting M81
group, could be tidal dwarf galaxies.

Table 2 summarizes the results of the
synthetic CMD modelling of our HST/WFPC2 data for the suspected tidal dwarfs.
The distances to the galaxies and Galactic foreground extinction values
used in the computation are given in columns 2--4
of Table~2.

\begin{table*}
\setcounter{table}{1}
\caption{Star formation history of the tidal dwarfs.}
\begin{tabular}{ccclllc}
\hline
Name        &$E(V-I)$&$A_I$& $\mu_0$& Star formation age & mean SFR  & [Fe/H] \\
	    & [mag]&[mag]&[mag]   &                    & [M$_{\sun}$~yr$^{-1}$]     & [dex]  \\
\hline
&&&&&&\\
Holmberg~IX & 0.109 & 0.154 &  27.80 & 6--200~Myr & 7.5$^{+2.6}_{-2.2}\cdot$10$^{-3}$ & --0.7$\div$--0.4 \\
&&&&&&\\
\hline
&&&&&&\\
BK3N        & 0.110 & 0.155 &  27.80 & 8--200~Myr & 3.8$^{+8.9}_{-1.7}\cdot$10$^{-4}$ & --0.7$\div$0.0 \\
&&&&&&\\
	    &&&& 500~Myr--2~Gyr & 0.8$^{+13.4}_{-0.5}\cdot$10$^{-4}$ & --0.7 \\
&&&&&&\\
	    &&&& 8--12~Gyr & 2.0$^{+2.9}_{-1.4}\cdot$10$^{-4}$ & --0.7 \\
&&&&&&\\
\hline
&&&&&&\\
Arp-loop    & 0.117 & 0.164 &  27.80 & 40~Myr--160~Myr & 3.6$^{+9.7}_{-2.1}\cdot$10$^{-4}$ & --1.3$\div$--0.4 \\
&&&&&&\\
	    &&&& 500~Myr--1~Gyr  & 2.6$^{+13.9}_{-1.4}\cdot$10$^{-4}$ & --1.3$\div$--0.4 \\
&&&&&&\\
	    &&&&  5--8~Gyr       & 2.2$^{+5.2}_{-1.6}\cdot$10$^{-4}$ & --1.3$\div$--0.7 \\
&&&&&&\\
\hline
&&&&&&\\
Garland     & 0.092 & 0.130 &  27.89 & 16--200~Myr~:    & & --0.7$\div$0.0 \\
&&&&&&\\
	    &&&& 600~Myr--1~Gyr~: & & --0.7$\div$0.0 \\
&&&&&&\\
	    &&&& 8--10~Gyr~:      & & --0.7$\div$0.0 \\
&&&&&&\\
\hline
\end{tabular}
\end{table*}

The photometric limits of our images do not allow us to clarify the 
evolutionary
status of the four dwarf irregular galaxies 
with good accuracy. However,
a comparison of the star formation ages in the galaxies derived via
synthetic CMD modelling helps us to make
some suggestions. As can be seen from Table~2, Holmberg IX, BK3N, and
Arp-loop (and probably Garland as well) experienced an episode of enhanced
star formation between roughly 20 and 200~Myr ago.  The star formation
process may have been activated by the tidal interaction between 
M81, M82, and NGC3077, which is believed to have 
occurred 220~Myr ago (M~81--M~82) and 280~Myr ago (M~81--NGC~3077) according
to numerical simulations by Yun (\cite{y99}).

There are also intermediate age stars in Arp-loop and
BK3N (and probably in Garland, too). These stars may indicate that these
two (three) galaxies existed already 
prior to the tidal interaction, or contain older stellar material torn out
of the massive galaxies during the interaction.  
Finally, some star formation may have occurred in Arp-loop and BK3N
$\ge$ 8~Gyr ago, but these older stars have magnitudes close to our
detection limit, which makes their interpretation difficult.
While the young populations show a clear spatial coincidence and 
concentration toward the centers of the dwarf galaxies, this is
less pronounced for the older populations.  The proposed old red giant
branch population may in fact belong to the extended old population of M~81
(and NGC~3077)
and could be seen in projection along the line of sight to the dwarfs.
In the case of Holmberg~IX no significant indications of star formation longer
ago than $\sim 200$~Myr ago could be detected.

Judging from the results of the synthetic CMD modelling, the 
metallicity of the detected stars shows a spread of up to 1 dex
and is relatively high (up to $-0.4$ dex in [Fe/H]).  While
the caveats of deriving photometric metallicities of young populations
via isochrones should be kept in mind, these relatively high values
(for low-mass dwarf galaxies) may support a tidal origin, although
they remain below the abundances measured in M~81 H\,{\sc ii} regions.

Considering the similarity of the star formation histories of the four
dwarfs, they may all have formed from material of the metal-poor outer
part of the giant spiral galaxy M~81 after the tidal interaction event
about 200 Myr ago.  The ultimate confirmation of this scenario will
require significantly deeper CMDs than our present HST data provide.

\begin{acknowledgements}
We are very grateful to S.~A.~Pustilnik and A.~G.~Pramsky for kindly
making available to us the results of their spectroscopic measurements.
LNM thanks J.\ Harris for his help with the program package starFISH.
The work of LNM was supported by INTAS grant YSF 2001/1-0129
and by the Max-Planck Institute for Astronomy, Heidelberg.
LNM, IDK and MES acknowledge support through the
Russian Foundation for Base Research Grant 01-16001.
D.G. gratefully acknowledges support from the Chilean
{\sl Centro de Astrof\'\i sica} FONDAP No. 15010003.
\end{acknowledgements}

{}

\newpage

\begin{figure*}[h]
\psfig{figure=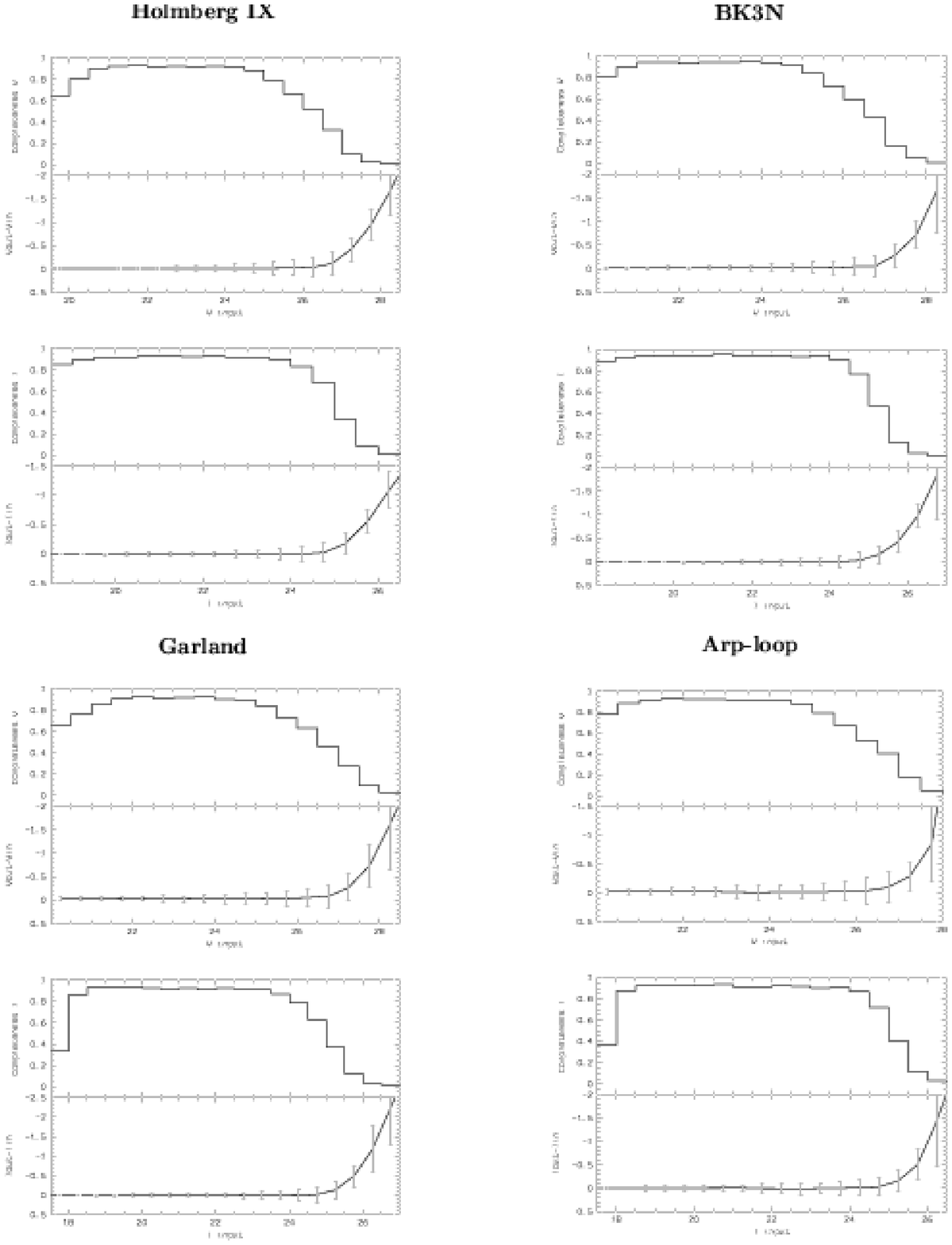,width=18cm}
\caption{V and I completeness (upper panel) and photometric errors
(bottom panel) for the
dwarf galaxies. In the bottom panel for each galaxy error bars show
the 1$\sigma$ distribution.}
\end{figure*}

\begin{figure*}[h]
\psfig{figure=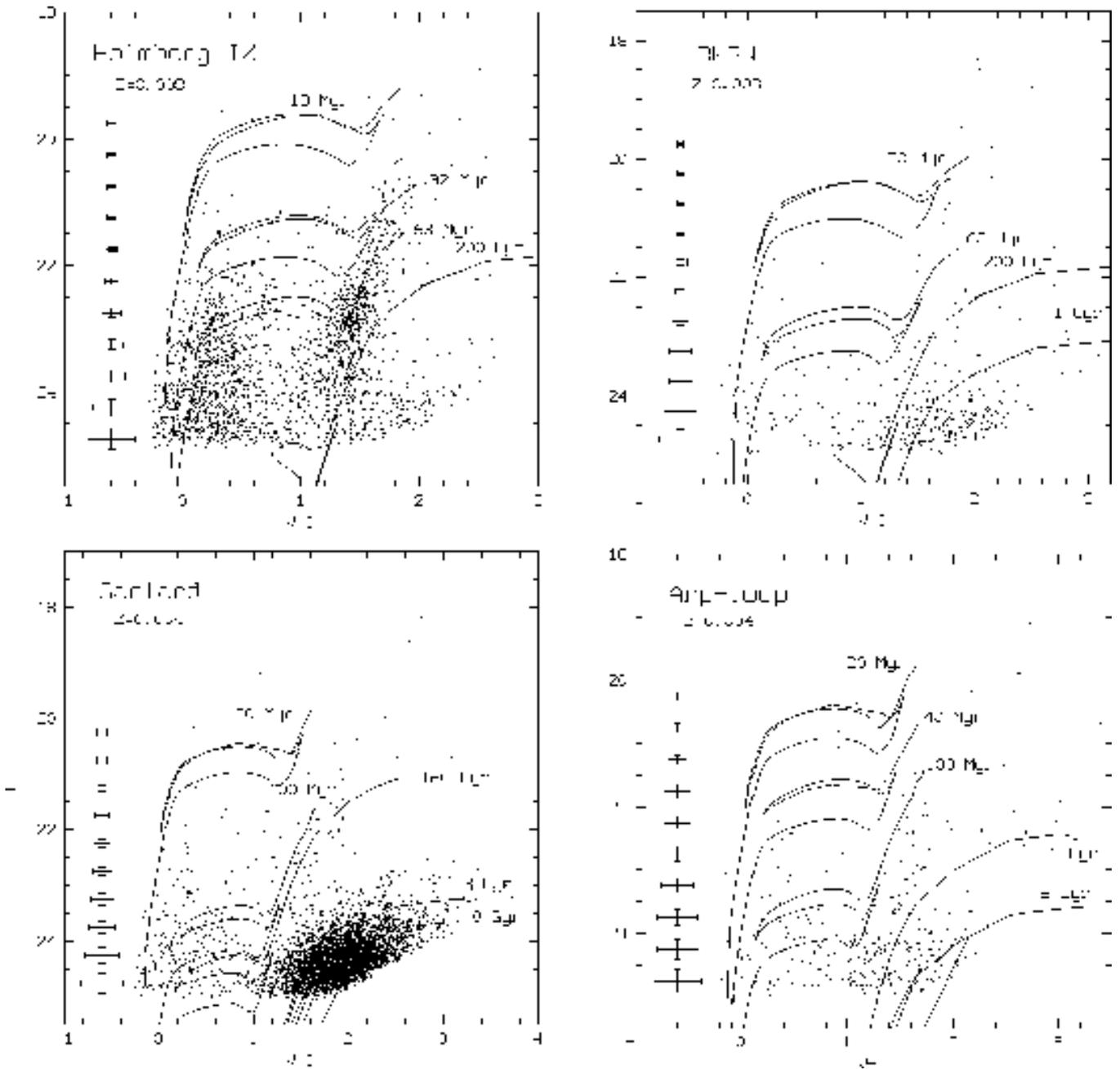,width=18cm}
\caption{I,V--I color-magnitude diagrams of the dwarf irregular galaxies.
The error bars on the left side of each panel show 1~$\sigma$ distribution
of the photometric errors. Isochrones by Girardi et~al.\ (\cite{girardi}) are
overplotted in the CMDs.}
\end{figure*}

\begin{figure*}[h]
\psfig{figure=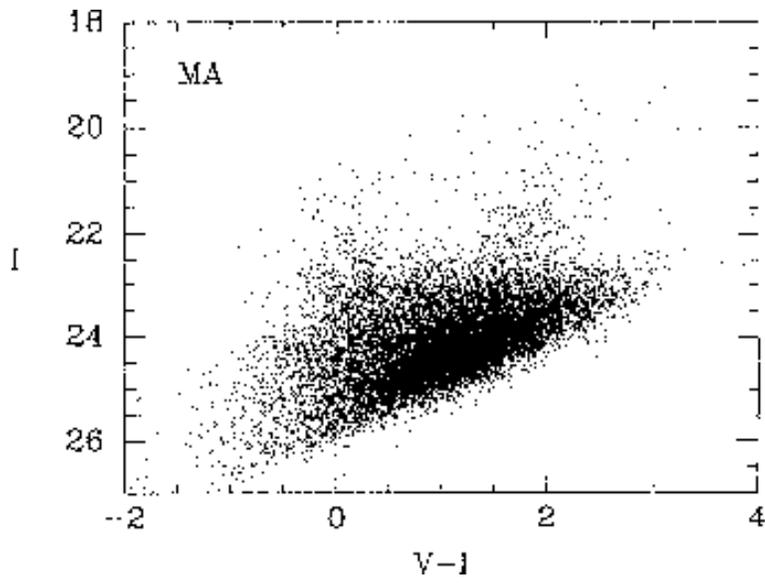,width=10cm}
\caption{CM diagram of M~81 major axis field from Hughes et~al.\ (\cite{hughes}).
Reproduced by permission of the AAS.}
\end{figure*}

\begin{figure*}[p]
\psfig{figure=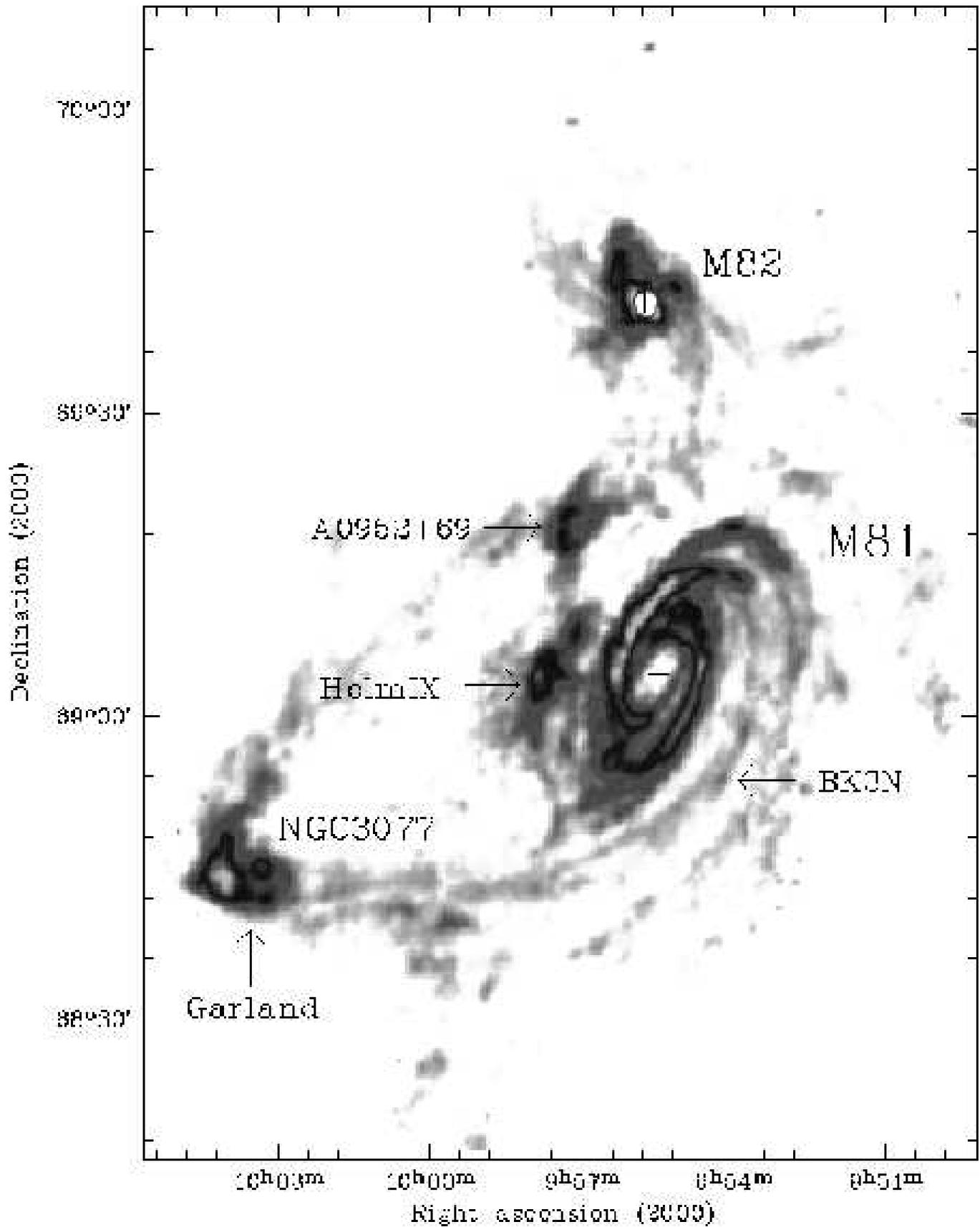,width=18cm}
\caption{Integrated HI map of the core M~81 group from Yun et~al.\ (\cite{yun}).
Reproduced by permission of the Nature.}
\end{figure*}

\begin{figure*}[p]
\bigskip
\psfig{figure=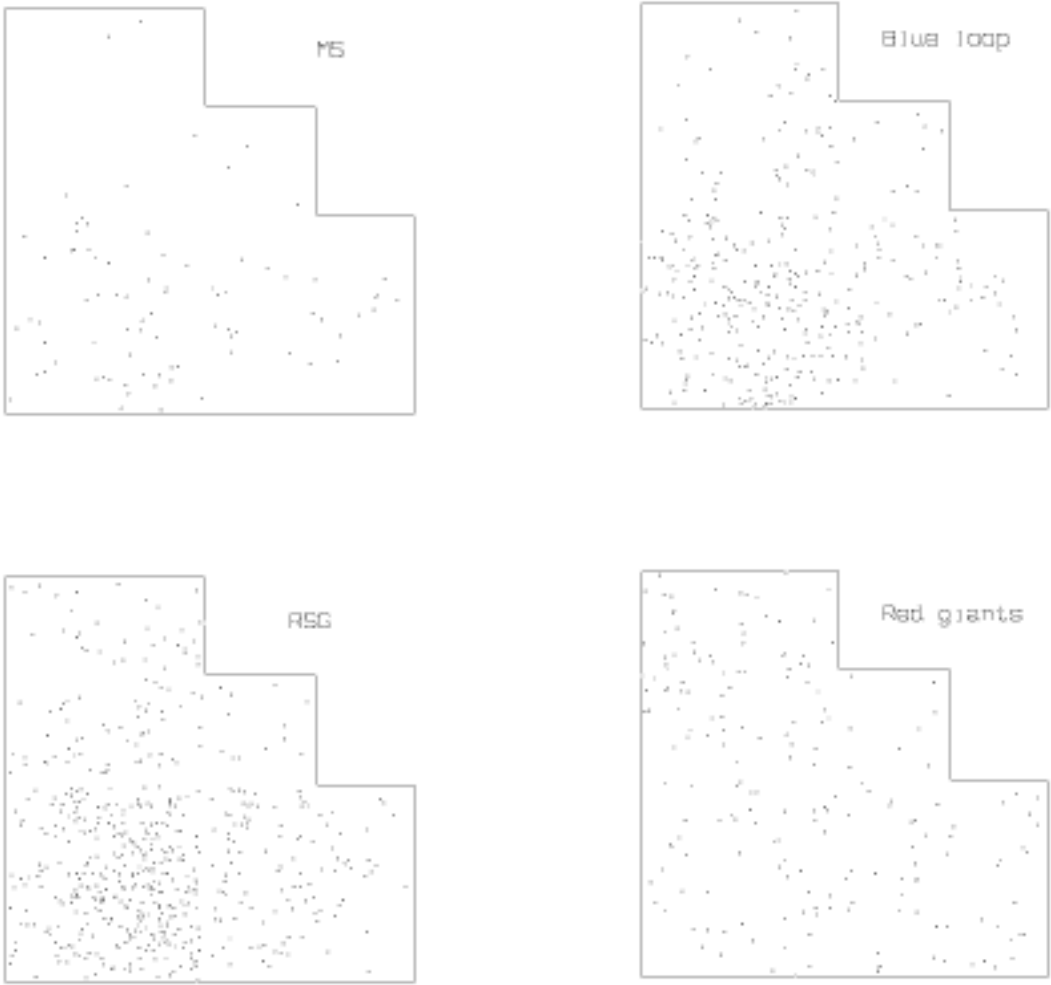,width=14cm}
\bigskip
\caption{Spatial distribution of the stellar populations in Holmberg~IX.
The main sequence stars (their age range is about 10--100~Myr) is bounded by
(V--I)$_0$~$<$~0 and I$_0$~$\le$~23.75.
The boundaries of the shown blue loop ($\sim$10--200~Myr) are~: 0~$\le$~(V--I)$_0$~$\le$~0.75 and
I$_0$~$\le$~23.75. RSG stars ($\sim$10~Myr--1~Gyr) correspond to the range~: (V--I)$_0$~$>$~0.75
and I$_0$~$\le$~23.75. The probable red giants ($>$1~Gyr) bounded by (V--I)$_0$~$\ge$~0.8
and 23.75~$\le$~I$_0$~$\le$~24.5.}
\end{figure*}

\begin{figure*}[h]
\vspace{3cm}
\psfig{figure=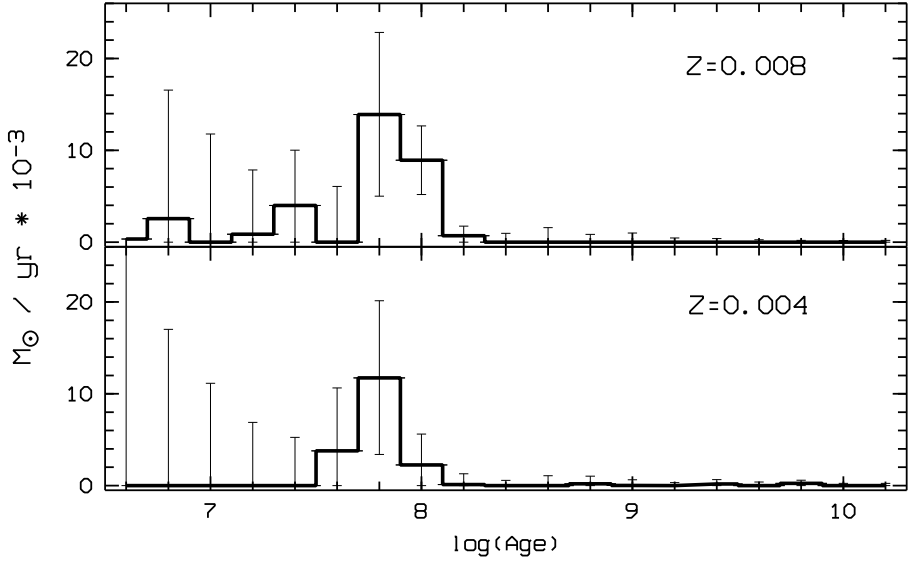,width=12cm,angle=-90}
\caption{The best-fit SFH for Holmberg IX. Error bars on the SFH
amplitudes are determined by identifying the 68~\% (1~$\sigma$) confidence
interval on each amplitude (see Harris \& Zaritsky (\cite{hz}) for details).
}
\end{figure*}

\begin{figure*}[h]
\vspace{3cm}
\psfig{figure=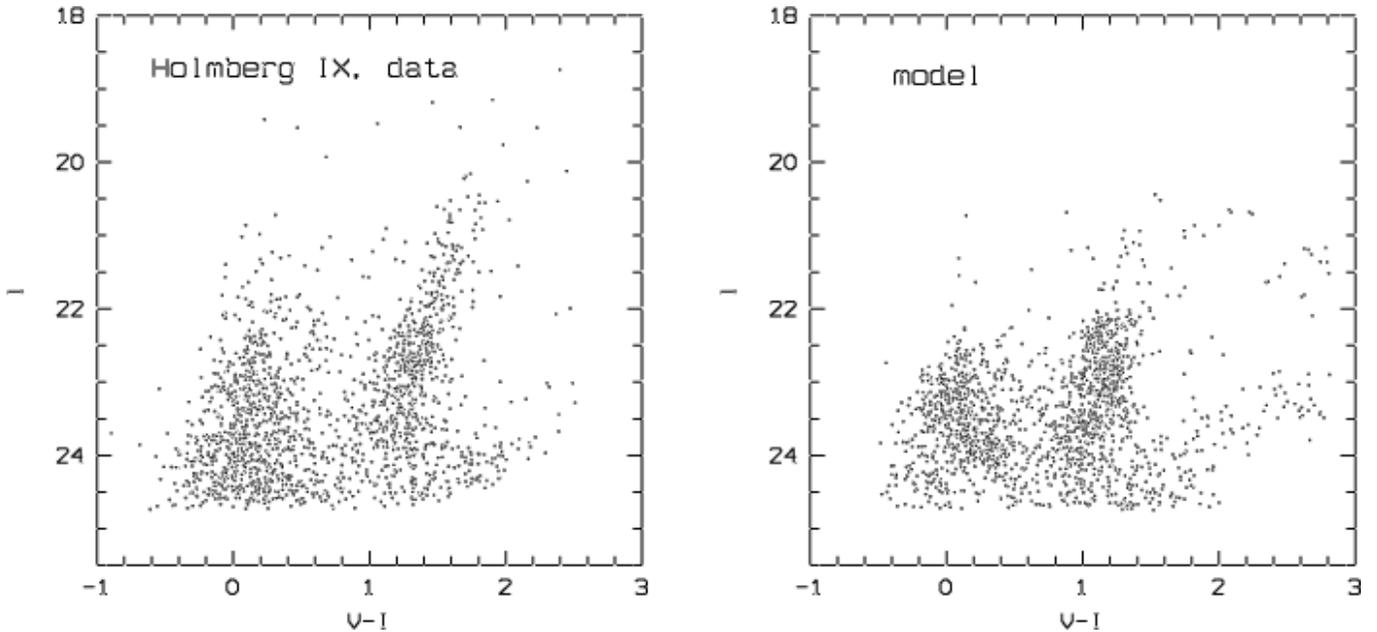,width=18cm,angle=-90}
\caption{The best-fit model CMD (right panel) and observed CMD
(left panel) of Holmberg~IX.}
\end{figure*}

\begin{figure*}[h]
\bigskip
\psfig{figure=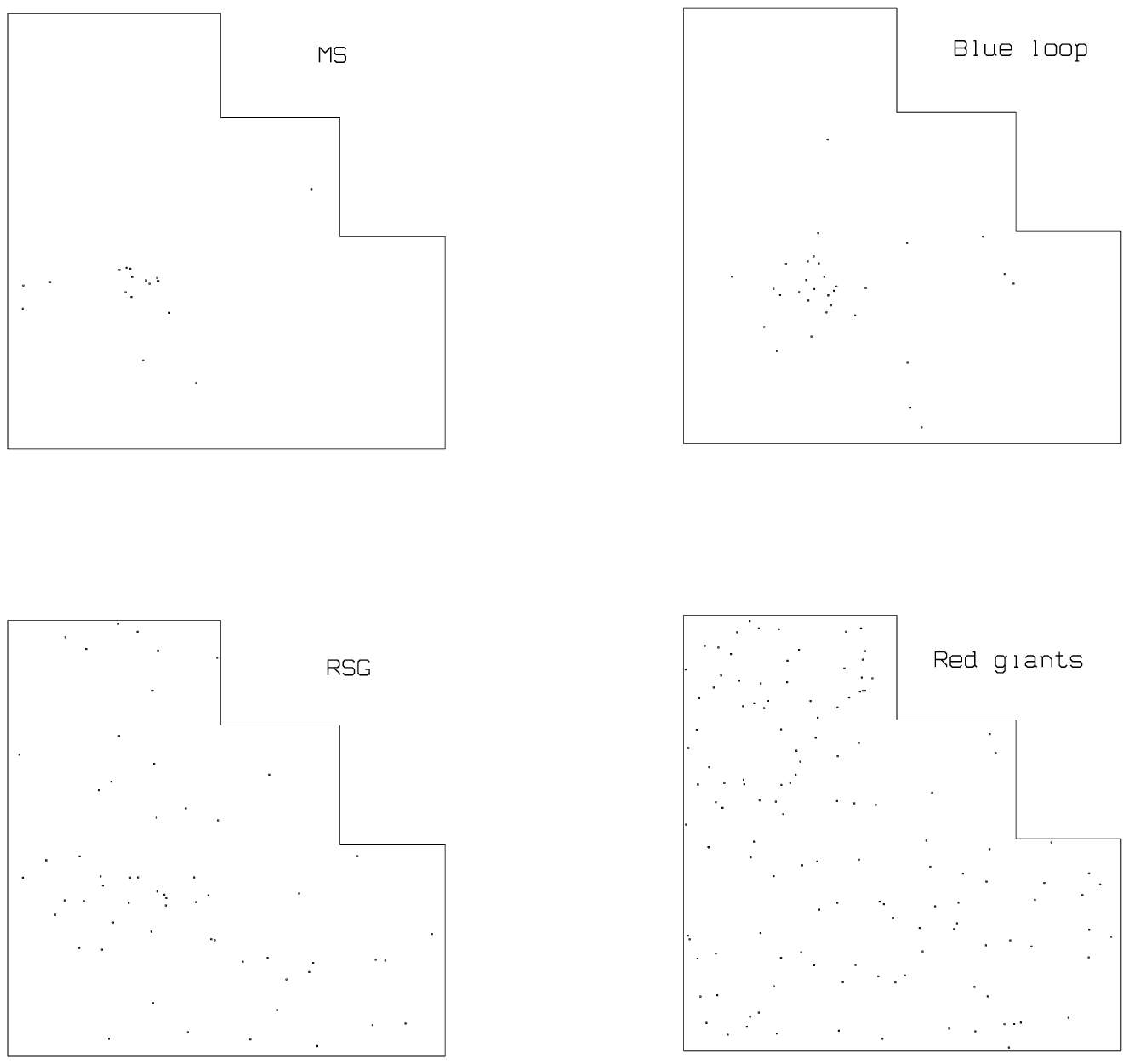,width=14cm}
\bigskip
\caption{Spatial distribution of the stellar populations in BK3N.
The main sequence stars ($\sim$10--100~Myr) is bounded by (V--I)$_0$~$<$~0 and I$_0$~$\le$~24.5.
The boundaries of the blue loop ($\sim$10--100~Myr) are~: 0~$\le$~(V--I)$_0$~$\le$~0.75 and
I$_0$~$\le$~24.5. RSG stars ($\sim$10~Myr--1~Gyr) correspond to the range~: (V--I)$_0$~$>$~0.75
and I$_0$~$\le$~23.75. The red giants ($>$1~Gyr) bounded by (V--I)$_0$~$\ge$~0.8
and 23.75~$\le$~I$_0$~$\le$~24.5.}
\end{figure*}

\begin{figure*}[h]
\vspace{1cm}
\psfig{figure=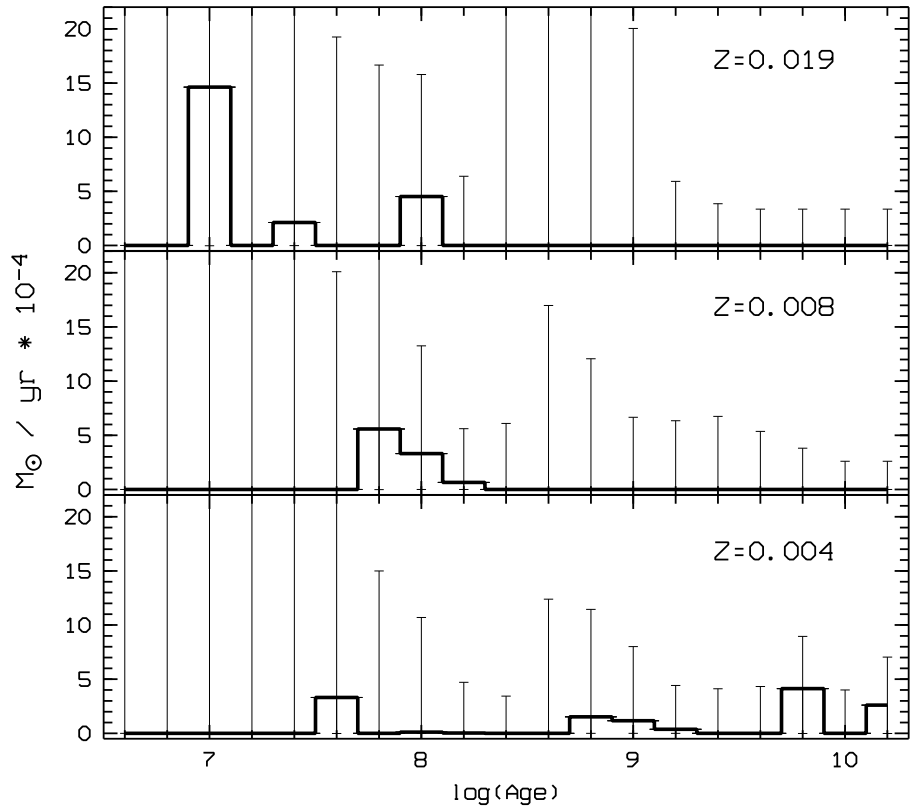,width=12cm,angle=-90}
\caption{The best-fit SFH for BK3N.}
\end{figure*}

\begin{figure*}[h]
\vspace{3cm}
\psfig{figure=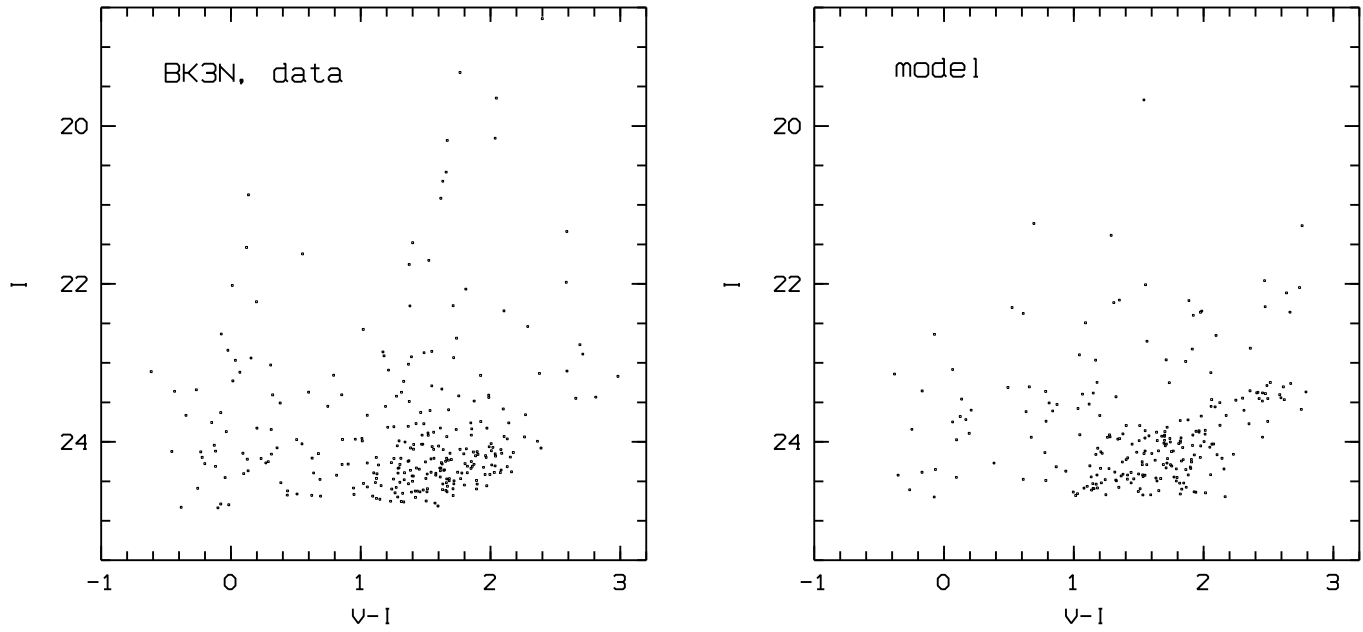,width=18cm,angle=-90}
\caption{The best-fit model CMD (right panel) and observed CMD
(left panel) of BK3N.}
\end{figure*}

\begin{figure*}[h]
\bigskip
\psfig{figure=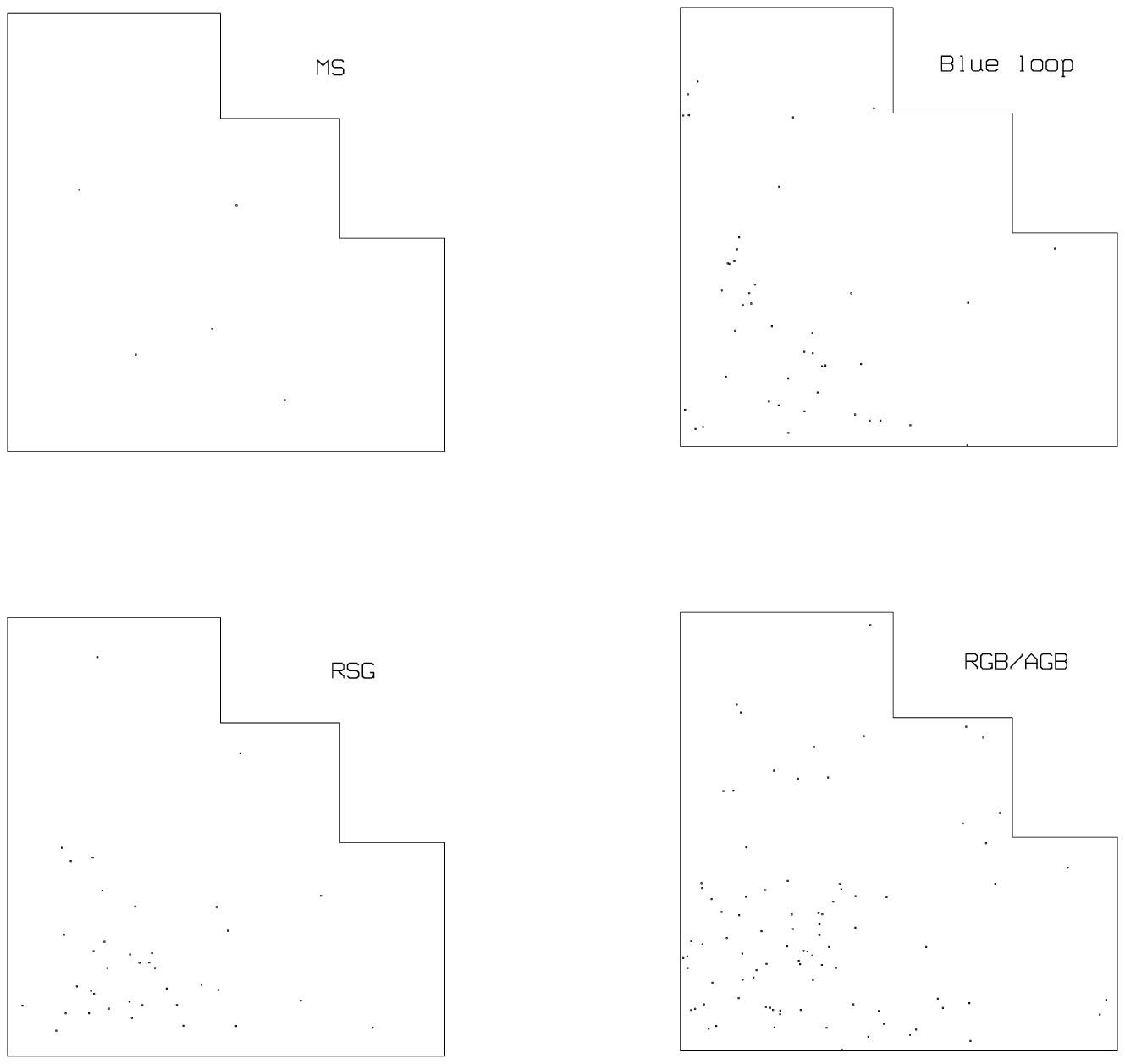,width=14cm}
\caption{Spatial distribution of the stellar populations in Arp-loop.
The main sequence stars ($\sim$40--200~Myr) is bounded by (V--I)$_0$~$<$~0 and I$_0$~$\le$~24.5.
The boundaries of the blue loop ($\sim$40--200~Myr) are~: 0~$\le$~(V--I)$_0$~$\le$~0.75 and
I$_0$~$\le$~24.5. RSG stars ($\sim$40~Myr--1~Gyr) correspond to the range~: 0.75~$<$~(V--I)$_0$~$\le$~2
and I$_0$~$\le$~23.75. The red giants ($>$1~Gyr) bounded by (V--I)$_0$~$\ge$~0.8
and 23.75~$\le$~I$_0$~$\le$~24.5.}
\end{figure*}

\begin{figure*}[h]
\vspace{3cm}
\psfig{figure=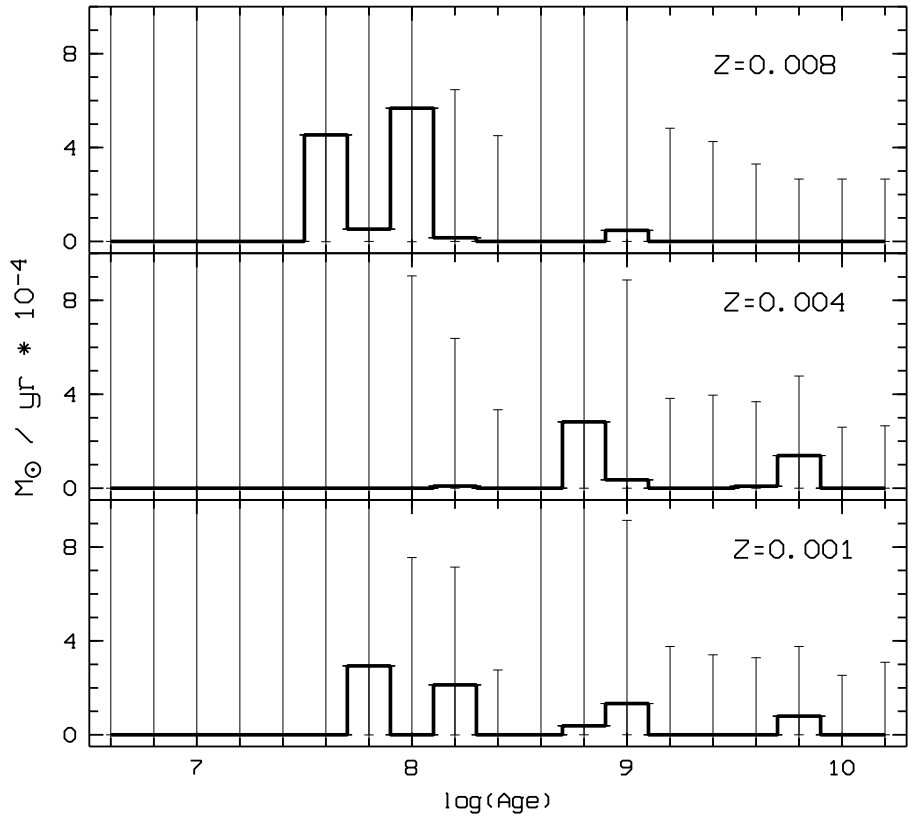,width=12cm,angle=-90}
\caption{The best-fit SFH for Arp-loop.}
\end{figure*}

\begin{figure*}[h]
\vspace{3.2cm}
\psfig{figure=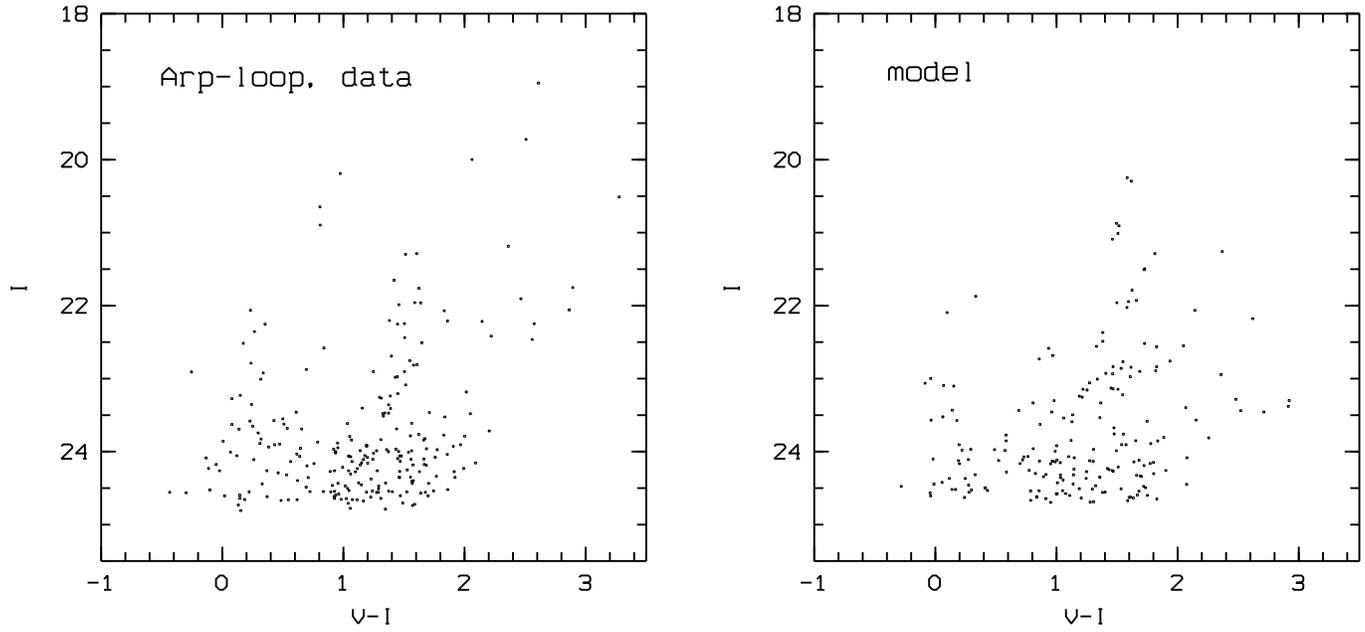,width=18cm,angle=-90}
\caption{The best-fit model CMD (right panel) and observed CMD
(left panel) of Arp-loop.}
\end{figure*}

\begin{figure*}[h]
\bigskip
\psfig{figure=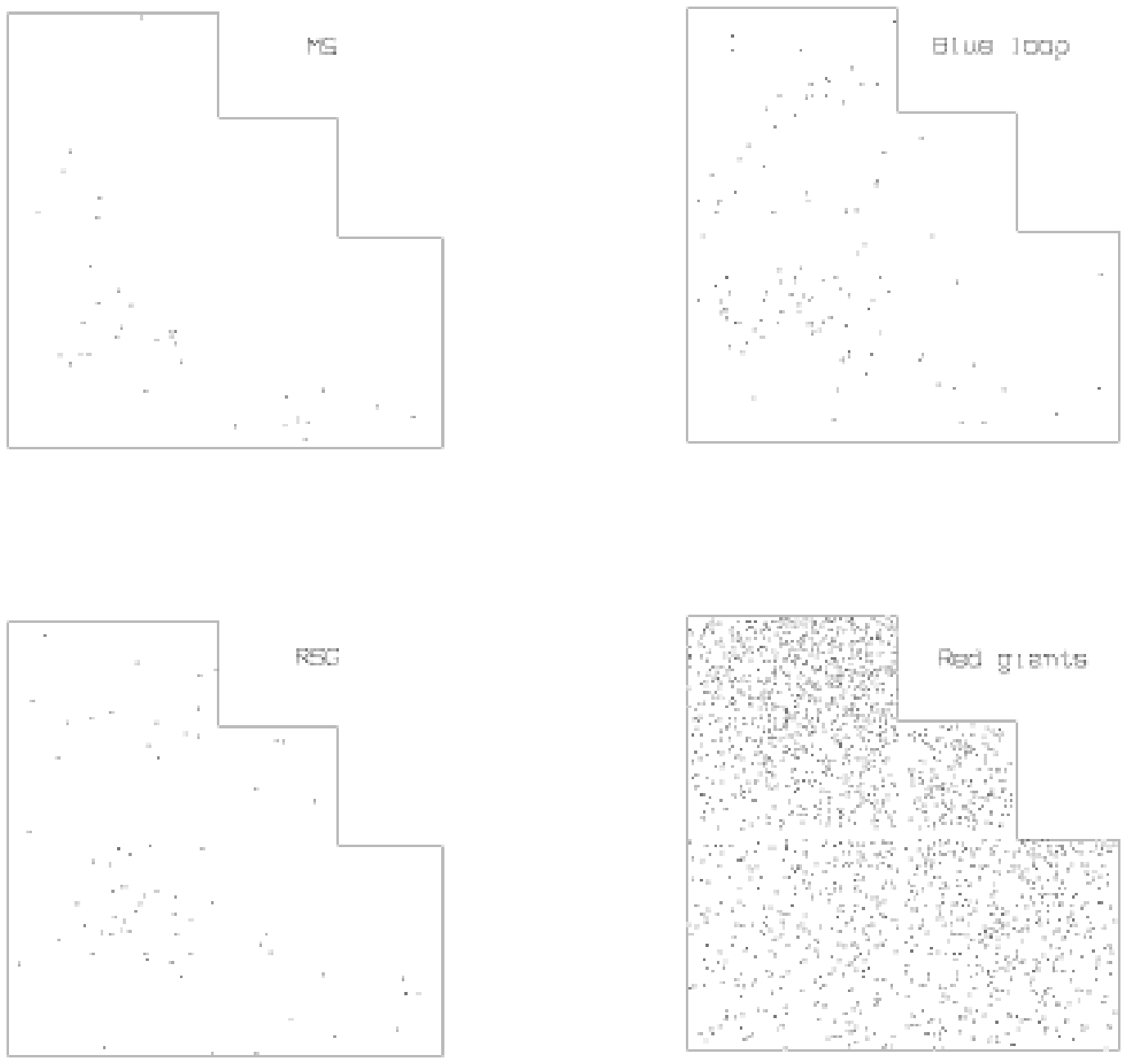,width=14cm}
\caption{Spatial distribution of the stellar populations in Garland.
The main sequence stars ($\sim$20--160~Myr) is bounded by (V--I)$_0$~$<$~0 and I$_0$~$\le$~24.5.
The boundaries of the blue loop ($\sim$20--160~Myr) are~: 0~$\le$~(V--I)$_0$~$\le$~0.75 and
I$_0$~$\le$~24.5. RSG stars ($\sim$20--160~Myr) correspond to the range~: 0.75~$<$~(V--I)$_0$~$\le$~2,
I$_0$~$\le$~23.75 and (V--I)$_0$~$>$~2, I$_0$~$\ge$~22. The red giants ($>$1~Gyr) bounded by
(V--I)$_0$~$\ge$~1
and 23.75~$\le$~I$_0$~$\le$~24.5.}
\end{figure*}


\begin{thebibliography}{}
\bibitem[1981]{appleton} Appleton P., Davies R., Stephenson R. 1981, MNRAS 195, 327
\bibitem[1981]{bs} Bahcall J.N., Soneira R.M. 1981, ApJS 47, 357
\bibitem[1994]{bertelli} Bertelli G., Bressan A., Chiosi C., Fagotto F.,
Nasi E. 1994, A\&AS 106, 275
\bibitem[1982]{bk} Boerngen, F., Karachentseva, V.E. 1982, AN 303, 189
\bibitem[2001]{boyce} Boyce P.J., Minchin R.F., Kilborn V.A. et~al.\ 2001,
ApJ, 560, L127
\bibitem[2002]{denicolo} Denicol\'o, G., Terlevich, R. \& Terlevich, E. 2002,
MNRAS, 330, 69
\bibitem[2000a]{dolphina} Dolphin A.E. 2000a, PASP 112, 1383
\bibitem[2000b]{dolphinb} Dolphin A.E. 2000b, PASP 112, 1397
\bibitem[2001]{d2001} Dolphin A.E., Makarova L., Karachentsev I.D. et~al.\
2001, MNRAS 324, 249
\bibitem[1994]{freedman} Freedman W.L., Hughes Sh.M., Madore B.F. et~al.\ 1994,
ApJ 427, 628
\bibitem[1994]{gsf} Gardiner, L.T., Sawa, T., Fujimoto, M. 1994,
MNRAS 266, 567
\bibitem[2000]{girardi} Girardi, L., Bressan, A., Bertelli, G., Chiosi, C.,
2000, A\&AS 141, 371
\bibitem[2000]{grebel} Grebel, E.K.,  Seitzer, P., Dolphin, A.E., Geisler, D.,
Guhathakurta, P.,
Hodge, P.W., Karachentsev, I.D., Karachentseva, V.E., Sarajedini, A.
2000, in Stars, Gas, and Dust in Galaxies: Exploring the Links,
ASP Conf. Ser. 221, eds.\ D.\ Alloin, K.\ Olsen, \& G.\ Galaz (San
Francisco: ASP), 147
\bibitem[2001]{Harbeck} Harbeck, D., Grebel, E.K., Holtzman, J., 
Guhathakurta, P., Brandner, W., Geisler, D., Sarajedini, A., Dolphin, A., 
Hurley-Keller, D., \& Mateo, M. 2001, AJ, 122, 3092
\bibitem[2001]{hz} Harris J., Zaritsky D. 2001, ApJ Suppl. 136, 25
\bibitem[2000]{hw} Heithausen, A., Walter, F. 2000, A\&A 361, 500
\bibitem[1994]{hughes} Hughes, S.M.G., Stetson, P.B., Turner, A.,
et al., 1994, ApJ 428, 143
\bibitem[2000]{hunter} Hunter D.A., Hunsberger S.D., Roye E.W. 2000,
ApJ 542, 137
\bibitem[1994]{ibata} Ibata, R.A., Gilmore, G., Irwin, M.J. 1994,
Nature 370, 194
\bibitem[1985]{kkb} Karachentseva, V.E., Karachentsev, I.D., Boerngen, F.
1985, A\&AS 60, 213
\bibitem[2001]{k2001} Karachentsev, I.D., Karachentseva, V.E., Huchtmeier,
W.K. 2001, A\&A, 366, 428
\bibitem[1999]{k99} Karachentsev I.D., Sharina M.E., Grebel E.K., Dolphin A.E.,
Geisler D., Guhathakurta P., Hodge P.W., Karachentseva V.E., Sarajedini A.,
Seitzer P. 1999, A\&A 352, 399
\bibitem[2000]{k00} Karachentsev I.D., Karachentseva V.E., Dolphin A.E.,
Geisler D., Grebel E.K., Guhathakurta P., Hodge P.W., Sarajedini A.,
Seitzer P., Sharina M.E. 2000, A\&A 363, 117
\bibitem[2001]{k01} Karachentsev I.D., Sharina M.E., Dolphin A.E.,
Geisler D., Grebel E.K., Guhathakurta P., Hodge P.W., Karachentseva V.E., Sarajedini A.,
Seitzer P. 2001, A\&A 375, 359
\bibitem[2002]{karachentsev} Karachentsev I.D., Dolphin A. E., Geisler D.,
Grebel E.K., Guhathakurta P., Hodge P.W., Karachentseva V.E., Sarajedini A.,
Seitzer P., Sharina M.E. 2002, A\&A 383, 125
\bibitem[1995]{LanMaed} Langer, N., \& Maeder, A. 1995, A\&A, 295, 685
\bibitem[1993]{lee} Lee M., Freedman W., Madore B. 1993, ApJ 417, 553
\bibitem[2002]{makarova} Makarova, L.N., Karachentsev, I.D., Grebel, E.K.,
Barsunova, O. Yu. 2002, A\&A 384, 72
\bibitem[1994]{mh} Miller, B.W., Hodge, P. 1994, ApJ 427, 656
\bibitem[2001]{sm} Sakai Sh., Madore B. 2001, ApJ 555, 280
\bibitem[1999]{seitzer} Seitzer P., Grebel E.K., Dolphin A.E., et~al.\ 1999,
AAS 195, 801
\bibitem[1998]{schlegel} Schlegel D.J., Finkbeiner D.P., Davis M. 1998, ApJ
500, 525
\bibitem[1977]{hulst} van der Hulst, J. M. 1977, Ph.D. thesis, Univ. Groningen
\bibitem[1998]{vandriel} van Driel W., Kraan-Korteweg R., Binggeli B.,
Huchtmeier W. 1998, A\&AS 127, 397
\bibitem[1998]{vanzee} van Zee, L., Salzer, J.J., Haynes, M.P., O'Donoghue, A.A.,
Balonek, T.J. 1998, AJ, 116, 2805
\bibitem[1994]{yun} Yun M., Ho P., Lo K. 1994, Nature 372, 530
\bibitem[1999]{y99} Yun M.S. 1999, IAU Symp. 186, 81
\end{thebibliography}
\end{document}